\documentclass[9pt,twocolumn,twoside]{osajnl}

\journal{ao} 

\setboolean{shortarticle}{false}

\title{Developement of a stabilized Fabry-P{\'e}rot etalon based calibrator for Hanle Echelle Spectrograph (HESP)}

\author[1,*]{Tanya Das}
\author[1]{Ravinder K. Banyal}
\author[1]{Sivarani T.}
\author[1]{Ravindra B.}

\affil[1]{Indian Institute of Astrophysics, Bangalore, India}

\affil[*]{Corresponding author: tanya@iiap.res.in}

\usepackage[utf8]{inputenc}
\usepackage{amsmath,amsfonts,amssymb}

\begin{abstract}
Accurate wavelength calibration is an important factor for any measurement with high resolution spectrographs. Stellar spectrum comprises of discrete absorption or emission lines whose position is precisely determined by calibrating the spectrograph using known reference lines generated from laboratory sources. For the spectrograph  to measure small variations in Doppler shift, the wavelength calibration must be sufficiently stable during observation time. Instrument instability, mainly due to environmental factors like temperature and pressure variations, limitations of traditional calibration methods, for example Th-Ar lamps, are the main challenges which high precision spectroscopy. Through proper environmental control, by maintaining pressure at few mbar and temperature fluctuations within $\pm$0.05~$^{\circ}${C}, Fabry P{\'e}rot etalon (FP) can yield a velocity precision of 1-10~m/s, when used for wavelength calibration. We have developed a passively stabilized FP based wavelength calibrator for Hanle Echelle Spectrograph (HESP) installed on Himalayan Chandra Telescope (HCT) at Indian Astronomical Observatory (IAO), Hanle, India. The etalon has been characterized using Fourier Transform Spectrograph (FTS) and initial test runs have been performed with HESP. In this paper we present the design and construction of the instrument along  with preliminary test results obtained from HESP.
\end{abstract}
\setboolean{displaycopyright}{true}
\begin{document}
\maketitle

\section{Introduction}
The detection of extrasolar planet orbiting  a main sequence star 51 Peg in 1995 using Doppler Spectroscopy or Radial Velocity (RV) technique, is one of the most significant discoveries made in astrophysics \cite{ref:mayor}. Since then, technological advancement has led to improved high precision photometric and spectroscopic techniques for detection of planets. To this date, the number of confirmed planets have crossed 4100 count from various ground based observations and space missions \cite{ref:nasa}. The RV technique is a powerful tool that contributed in many exoplanet discoveries before transit data from NASA's Kepler mission and later from Transiting Exoplanet Survey Satellite (TESS) became available. High precision radial velocity technique measures the Doppler shifts in the spectrum of the star. Apart from detection of extra solar planets, this technique finds application in areas such as, detection of g-mode oscillations in the Sun \cite{rf:appourchaux}, investigating expansion rate of the universe \cite{rf:liske}, measuring variability of fundamental constants \cite{rf:reinhold}, study of stellar interiors using astroseismology \cite{rf:bazot} and characterization of binary and pulsating stars \cite{rf:butler}.

In a gravitationally bound system, the star-planet pair orbits their common center-of-mass. Hence, the star is not completely stationary and appears to wobble slightly, causing a wavelength shifts in the spectrum. The spectrum appears red or blue shifted depending upon the movement of star away from or towards us. If there are periodic wavelength shifts in the star's spectrum, it ascertains the presence of invisible companion around the star. The amplitude of the wavelength shift depends on the mass and distance of the planet from the host star and the inclination angle that orbit makes with observer's line of sight. In case of binary systems, due to large mass of orbiting stars, the radial velocity variation of several kilometers per second are easily observed. Doppler shifts can be converted into radial velocity  using the following expression

\begin{equation}
 \frac{V_{RV}}{c} = \frac{\Delta\lambda}{\lambda}  \, ,
\label{eq:cnv}
\end{equation}   
where, $V_{RV}$ is the radial velocity, \textit{c} is speed of light, $\lambda$ is the wavelength in the rest frame and $\Delta\lambda$ is the change in wavelength due to Doppler motion. The RV signal is usually small due to high mass ratio between the star and the planet. For example, a  Jupiter-size planet would induce a radial velocity shift of 11.2~m/s in the spectra of a sun-like star from an orbital distance of 5.4~AU, while an Earth-sized object at 1~AU would induce a shift of $\sim$10~cm/s \cite{rf:hatzes}. Highly stable spectrographs dedicated to planet search programs have been able to reach a Doppler precision of 1~m/s, allowing detection of super earths ($5-8~M_{\oplus}$) \cite{rf:dittmann}.

Charge coupled device (CCD) is used for recording the spectra of star after dispersion from spectrograph grating. However, the CCD doesn’t record the wavelength but only the intensity information. Stellar spectrum consists of series of discrete absorption and emission lines produced due to a variety of chemical compositions and physical conditions prevailing in the star. The wavelength position of these lines can be precisely determined by referencing the spectrograph with known lines generated from laboratory sources like hollow cathode lamps \cite{rf:bauer}. The spectral intensity is recorded as a function of pixel position on a 2D grid. Grating being a non-linear device requires correct wavelength to be assigned to each pixel position on the CCD grid and hence the calibration is done by using a well characterized laboratory source. An empirical relation, known as wavelength solution, is used to map a discrete wavelength to each pixel position. This process is called wavelength calibration, whose accuracy is of paramount importance for any measurement with high resolution spectrographs.

The wavelength calibration must be sufficiently stable during observation time to measure small variations in Doppler shift. Instrument instability, induced by environmental factors like temperature and pressure variations, photon noise and limitations of traditional calibration methods are the major challenges to high precision spectroscopy \cite{rf:murphy}. Environmental fluctuations can easily induce errors of few 10-100~m/s in the spectrograph data. Even if the environmental variations are controlled, wavelength calibration errors would still remain \cite{rf:halverson}. Preceding discussions suggest that a meaningful comparison of the spectra taken at different times is often a difficult and nontrivial task. These challenges have inspired researchers to build highly stabilized spectrographs and explore innovative wavelength calibration techniques. Emission lamps, Iodine cells \cite{rf:marcy}, Laser Frequency Combs (LFC) \cite{rf:murphy} and Fabry-P{\'e}rot (FP) etalons \cite{rf:wildi} are the methods that are used for wavelength calibration. Emission lamps are the traditionally used technique of wavelength calibration, Thorium-Argon lamp being the most widely used. The lines provided by Th-Ar lamps are not uniform in distribution and intensity and suffers from blending caused by unresolved lines. Determining the line position in Th-Ar spectrum has an uncertainty of around 10~m/s \cite{rf:lovis}. When used in simultaneous calibration, the RV precision achievable with emission lamps can be improved. Iodine absorption cells provide a narrow wavelength coverage (500-630~nm), have high light losses and requires a complex code  to retrieve small Doppler shift. LFC is promising technology but comes with high costs and many operational difficulties. Providing stable spectrum for referencing with nearly equispaced lines of uniform intensity and easily matched wavelength range of the spectrograph, FP etalon can give frequency comb-like properties at affordable costs and less complexity.

A Fabry-P{\'e}rot etalon is a simple optical component consisting of two glass plates mounted parallel to each other, separated by a certain distance, with inner surfaces having high reflection coating \cite{rf:born}. The plates are wedged to avoid reflection from outer surfaces. The two glass plates form a resonant cavity and produce a periodically varying transmission function due to interfering multiple reflections of light in the cavity. In this paper, we present the design of a passively stabilized FP for tracking and calibration of Hanle Echelle Spectrograph (HESP) part of India's 2m Himalayan Chandra Telescope (HCT), to achieve an RV precision about 1-10~m/s. Temperature, pressure and humidity affect the performance of the etalon, which can be eliminated by installing the FP in a vacuum chamber that is temperature controlled. Section \ref{sec:stability} gives an overview of the design requirements of our instrument. In section \ref{sec:design} the various aspects of the FP instrument setup are presented. The transmission spectra of the FP etalon taken with a high resolution Fourier Transform Spectrograph (FTS) and test results from HESP are presented in section \ref{sec:test}. Finally, summary of the work is given in section \ref{sec:summary}.

\section{Factors Affecting Fabry-P{\'e}rot Stability}
\label{sec:stability}

The FP transmission function is produced by interference between the multiple reflections of light within the high reflecting surfaces of the FP. The transmission function of FP is given as \cite{rf:born}:
\begin{equation}
I_T = \frac{I_0}{1+Fsin^2(\delta/2)}
\label{eq:int}
\end{equation}

where,
\begin{equation}
F = \frac{\pi\sqrt{R}}{1-R}
\label{eq:fin}
\end{equation}

and
\begin{equation}
\delta = \frac{4\pi nLcos(\theta)}{\lambda}
\label{eq:del}
\end{equation}

Here, \textit{F} is coefficient of finesse, \textit{R} is mirror reflectivity, \textit{n} is refractive index of cavity, \textit{$\theta$} is angle of incidence on second surface and \textit{L} is the cavity length.

Free spectral range (FSR) is the separation between adjacent transmission peaks in the FP spectra. FSR in terms of wavelength ($\lambda$) can be expressed as:

\begin{equation}
FSR = \Delta\lambda = \frac{\lambda^2}{2nL}
\label{eq:fsr}
\end{equation}

\begin{figure}[htbp]
\centering
\vbox{\includegraphics[width=\linewidth]{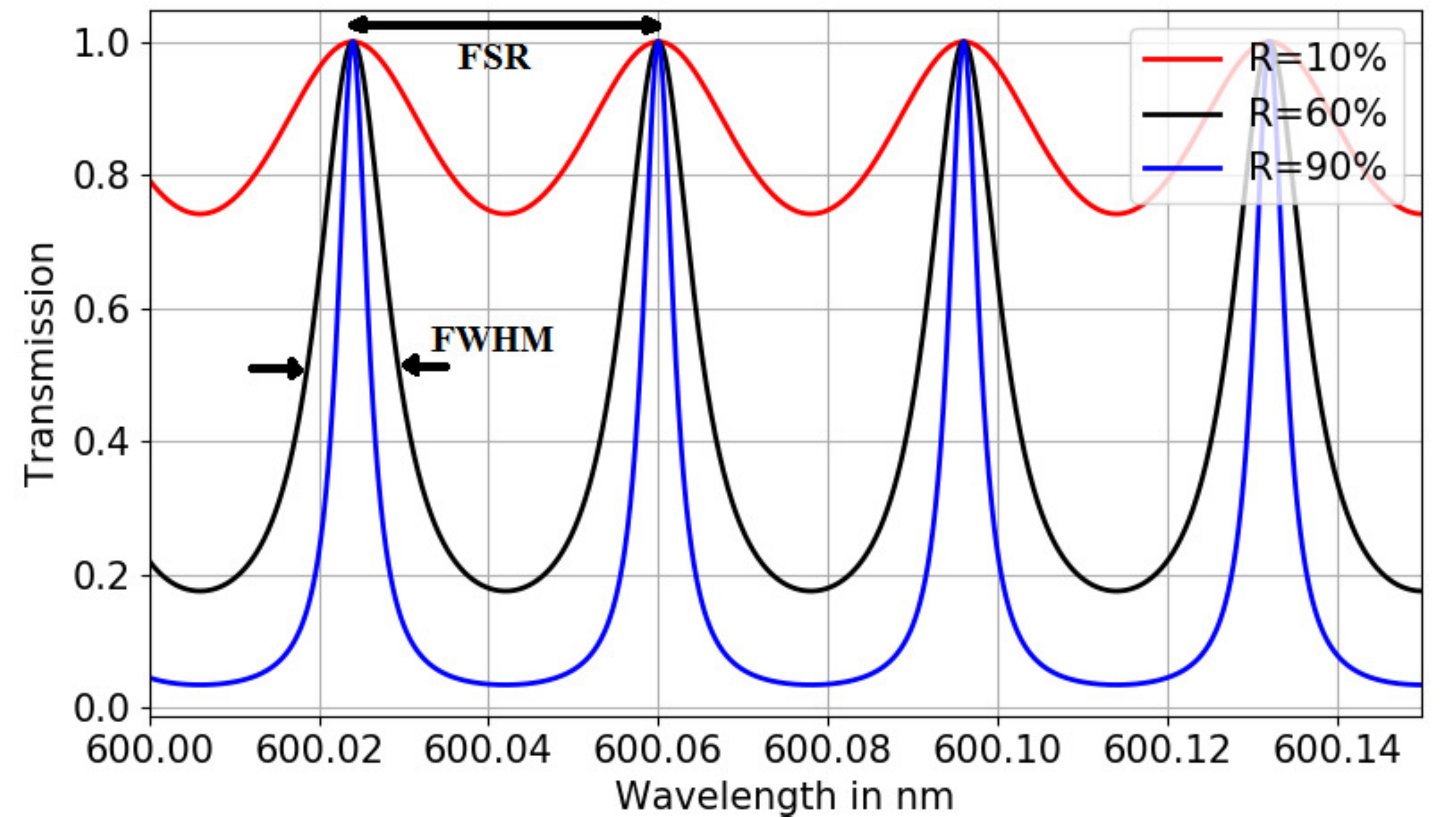}}
\caption{Simulated curve of FP transmission function for different values of reflectivity coating R. As the R increases, the FP transmission peaks become sharper. FSR is the free spectral range of the FP and FWHM is the line width.}
\label{fig:fp3r}
\end{figure}

Figure~\ref{fig:fp3r} shows the transmission spectra of FP with FSR and full width at half maximum (FWHM).

From Eq.~ \ref{eq:fsr}, it can be seen that changes in cavity length or refractive index of the cavity would change the FSR and cause random drifts in calibration lines. To ensure that the FP calibration lines remain stable, fluctuations in $n$ and $L$ have to be minimized. Getting stable transmission output from the FP depends upon the following factors:

\subsection{Pressure stability of the FP housing environment}
Refractive index of air is a function of temperature and pressure. Dispersion formula of dry air, as discussed in detail by  J. C. Owens in Ref.\cite{rf:owens},   is used to calculate the refractive index fluctuation for varying temperature and pressure. The refractive index variation is converted into RV shift using Eq.~ \ref{refr}. 

\begin{equation}
\frac{\Delta n}{n} = \frac{V_{RV}}{c} \, ,
\label{refr}
\end{equation}   
where, $V_{RV}$ is the radial velocity and \textit{c} is speed of light.

At atmospheric pressure, small changes in temperature strongly effect the refractive index of air, causing large errors in RV measurement. At low pressure, the effect of temperature variation on RV becomes less significant. Hence the FP setup is installed in an evacuated enclosure operating at pressure below 1~mbar. A detailed analysis has been presented in \cite{rf:das}.

\subsection{Temperature stability of the FP housing environment}
Fluctuations in ambient temperature can alter the FP cavity length. A spacer of very low thermal expansion material, like Zerodur and ULE, holds the glass plates of FP together. Shifts in RV measurement can be caused due to small variation in temperature induced cavity length change. The FP is being operated at 4 – 5~$^{\circ}${C} above the ambient temperatures and the temperature fluctuations are limited within 0.05~$^{\circ}${C} to obtain better RV precision. More implementation details can be found in Ref.~\cite{rf:das}.

\subsection{Misalignment errors}
\begin{figure}[htbp]
\centering
\vbox{\includegraphics[width= 8cm, height = 8 cm,keepaspectratio]{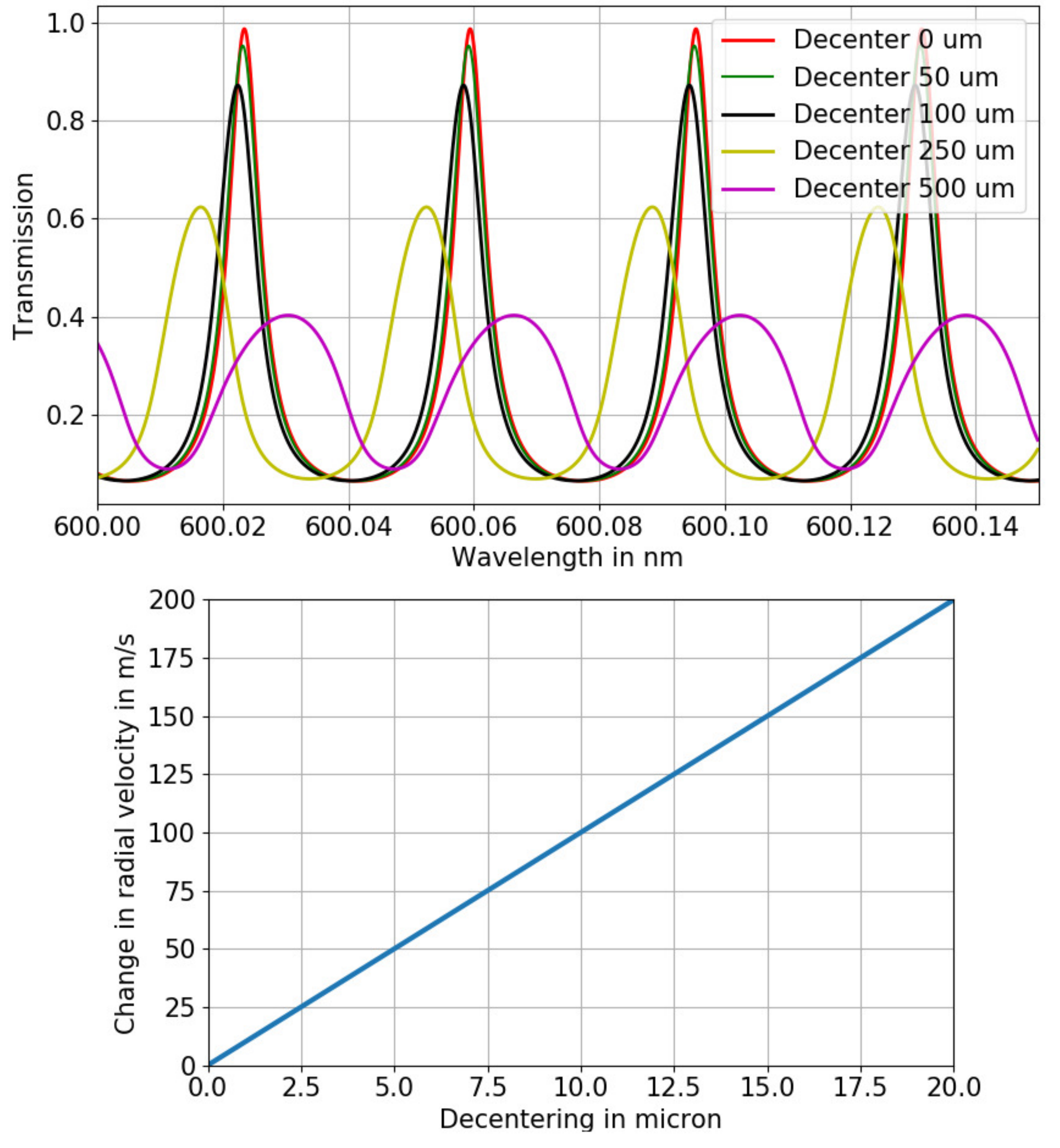}}
\caption{Top: Simulated transmission profile of FP for different fiber decentering values. Red curve shows the ideal FP output when there is no misalignment between the optical fiber and input optics. As the misalignment in the fiber increases, the quality of FP lines degrade as shown by the purple and green curves. Bottom: Radial Velocity error induced due to decentering of fiber. This sets a realistic limit on the misalignment that can be tolerated for the targeted RV precision.}
\label{fig:dec}
\end{figure}

\begin{figure}[htbp]
\centering
\vbox{\includegraphics[width= 8cm, height = 8 cm,keepaspectratio]{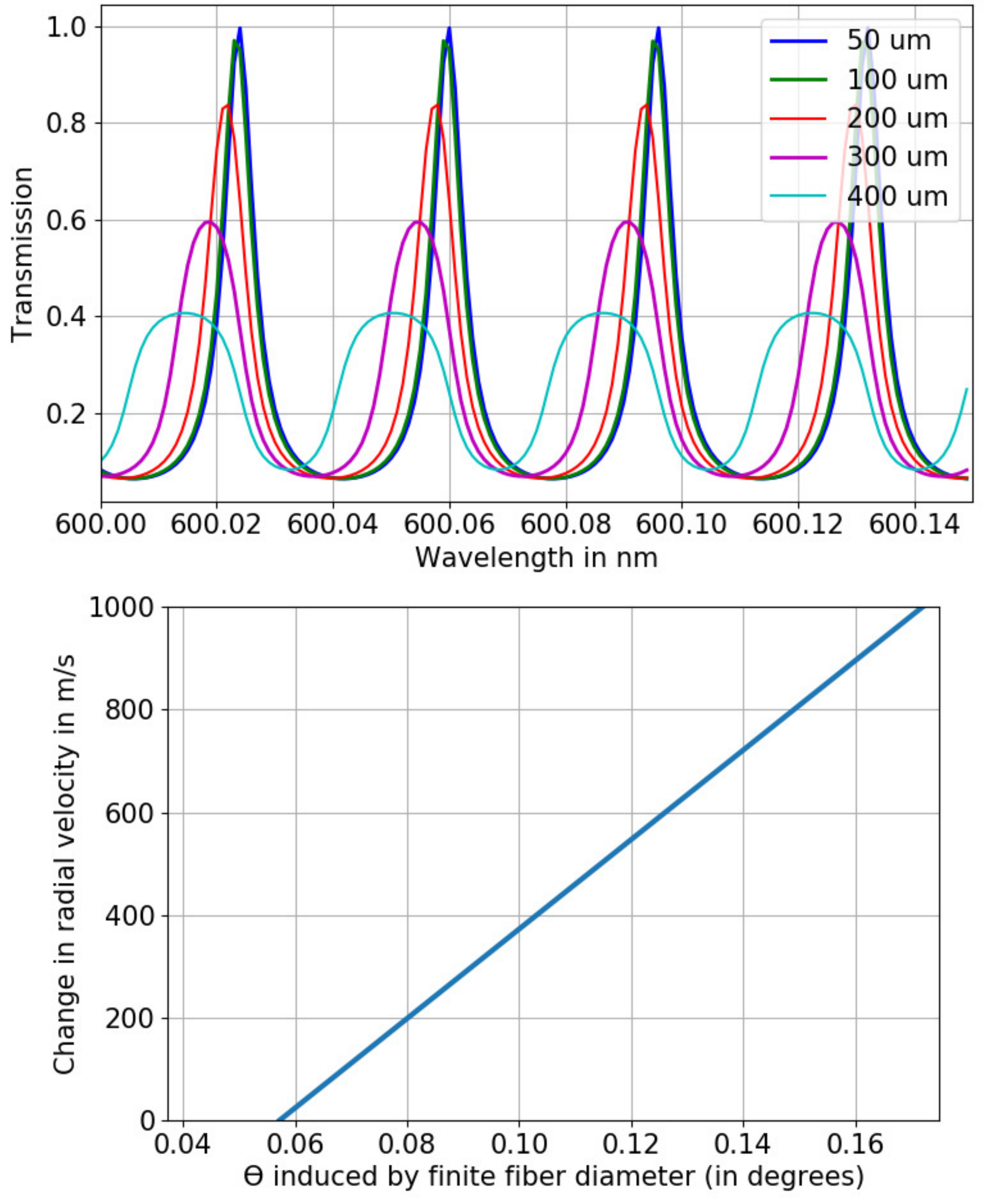}}
\caption{Top: Simulated transmission profile of FP illuminated with fibers of different sizes. As the fiber diameter increases, the FP line quality shows a progressive degradation. A larger core of the fiber acts as an extended source and hence leads to imperfect collimation. Bottom: Radial Velocity error induced in spectra due to finite divergence of fiber.}
\label{fig:dia}
\end{figure}

Effective finesse of the etalon depends upon reflectivity finesse, contribution from surface finish, parallelism finesse and imperfect collimation due to finite size of the light source \cite{rf:cersullo}. The beam divergence resulting from the finite size of the fiber core introduces loss in effective finesse and hence a shift in transmitted peaks towards the shorter wavelengths. Since the transmission function of the FP also depends upon the angle at which light is incident on the FP surface, it is important to ensure that FP is illuminated uniformly. Misalignment or decentering of the fiber will cause line broadening and loss in effective finesse.  Using the method discussed by F. Cersullo et al. in \cite{rf:cersullo}, we have simulated the decentering and effect of fiber size for the FP we are using and the RV error induced by each, as shown in Fig.~\ref{fig:dec} and Fig.~\ref{fig:dia}.

Based on the above considerations, the design requirements of the instrument are listed in Table~\ref{tab:desg}.

\begin{table}[htbp]
\centering
\caption{\bf Design requirements to achieve an RV precision about 10~m/s or better.}
\begin{tabular}{cc}
\hline
Design parameters & Required value  \\
\hline
Pressure & $\leq$1 mbar  \\
Temperature stability & $\pm$0.05 $^{\circ}${C}  \\
Misalignment error & $<$ 5 $\mu$m \\
Fiber size & $\leq 50~\mu$m \\
\hline
\end{tabular}
\label{tab:desg}
\end{table}

\subsection{Coating dispersion}
\label{sec:coating}
For absolute wavelength calibration, the effective cavity spacing of etalon should be known with an accuracy better than $\Delta$~$\lambda$ / $\lambda \simeq 3 \times 10^{-8}$ for RV precision of 10~m/s. This implies that the cavity width should be known with an accuracy of 1.5~{\AA}. However, in practice the cavity mirror separation cannot be measured with an accuracy better than $\sim 1~\mu$m \cite{rf:bauer}. Determining the effective cavity width of the etalon and ensuring a constant FSR is further compromised by finite thickness of multilayer dielectric coating of the cavity mirrors. Photons of different wavelengths penetrate to different optical depths, leading to wavelength dependency of cavity spacing \textit{d($\lambda$)}. Therefore, the FP cavity lines, unlike LFC, will not be exactly periodic. This is one of the factors that could limit the use of FP for absolute wavelength calibration.\textbf{The functional dependence of coating dispersion can be predicted and corrected, provided the coating structure is known \cite{rf:mccr}. It is also possible to model the slow variations in FSR from the FP data taken with a high-resolution spectrograph.} For example, Bauer et al. \cite{rf:bauer} used a spline function to extract out  the contribution of wavelength dependent undulations of FSR. Other direct approach is to externally cross-calibrate the etalon with primary sources such as Fourier Transform Spectrograph or multiline Th-Ar lamp \cite{rf:wildi}. Once a unique wavelength is assigned to each FP peak, a secondary wavelength solution, without the need to consider coating dispersion, can be easily derived. Yet another way to ensure a constant FSR across the wavelength range is to use  dispersion-free  cavity mirrors \cite{rf:chen,rf:ma}

\section{Instrument Design}
\label{sec:design}

\begin{figure}[htbp]
\centering
\vbox{\includegraphics[width=\linewidth]{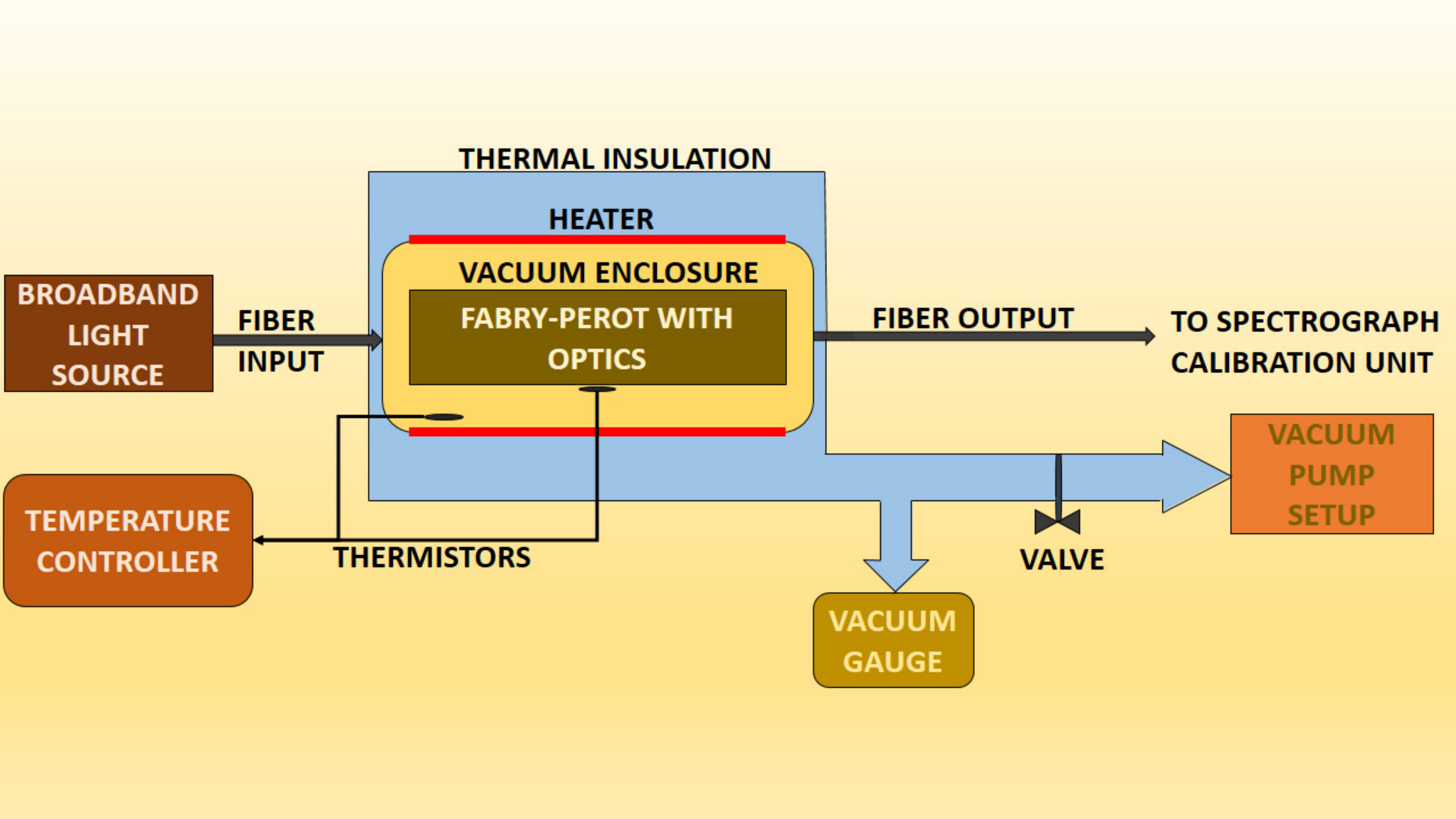}}
\caption{Schematic for complete set up of the Fabry-P{\'e}rot based wavelength calibrator.}
\label{fig:stp}
\end{figure}

Figure \ref{fig:stp} gives an overview of the instrument at component level. The instrument design comprises of two aspects, the optomechanical design and the design for environmental stability, namely, pressure and temperature control. The optomechanical design consists of proper mounting scheme for Fabry-Perot unit, beam shaping optics and fibers for light delivery and collection. The second aspect deals with the vacuum enclosure with relevant feedthroughs and pumps, and the temperature controller with temperature sensors.

\subsection{Pressure control}
FP unit is being housed in a standard stainless steel coupler of 333~mm length and 147~mm diameter. Fiber feedthroughs are used to provide access to the optical fibers into and out of the vacuum chamber. There are two types of feedthroughs available: penetrating-type; which have fibers directly running through the flange and receptacle-type; which uses two separate fibers on both side of vacuum flange. To avoid the light loss which occurs at the connection points between the fibers and feedthrough in receptacle type, we have used a penetrating type feedthrough fabricated in house using a method described in Ref.~\cite{ref:kirilov}.

\begin{figure}[htbp]
\centering
\vbox{\includegraphics[width=\linewidth]{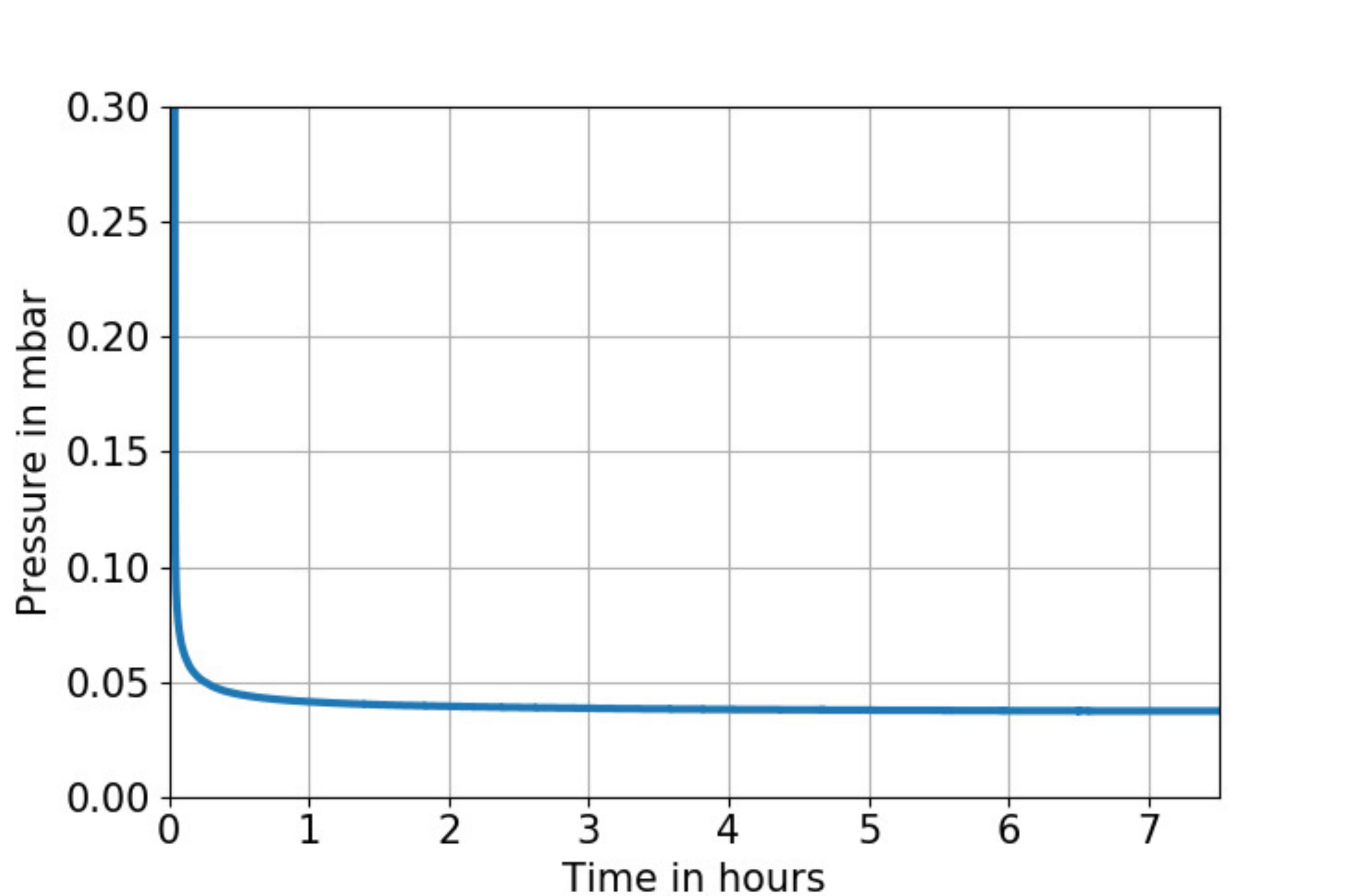}}
\caption{Pressure stability of the FP enclosure achieved during a laboratory test run.}
\label{fig:pres}
\end{figure}

Pressure stability of the setup has been tested in the laboratory. A vacuum pressure of 0.03~mbar was achieved in laboratory tests. A combination of roughing pump and turbo pump are being used to reach the desired vacuum level. The data for pressure has been logged for a period of 8 hours in the laboratory test and its plot is shown in Fig.~\ref{fig:pres}.

\subsection{Temperature control}
\begin{figure}[htbp]
\centering
\vbox{\includegraphics[width=\linewidth]{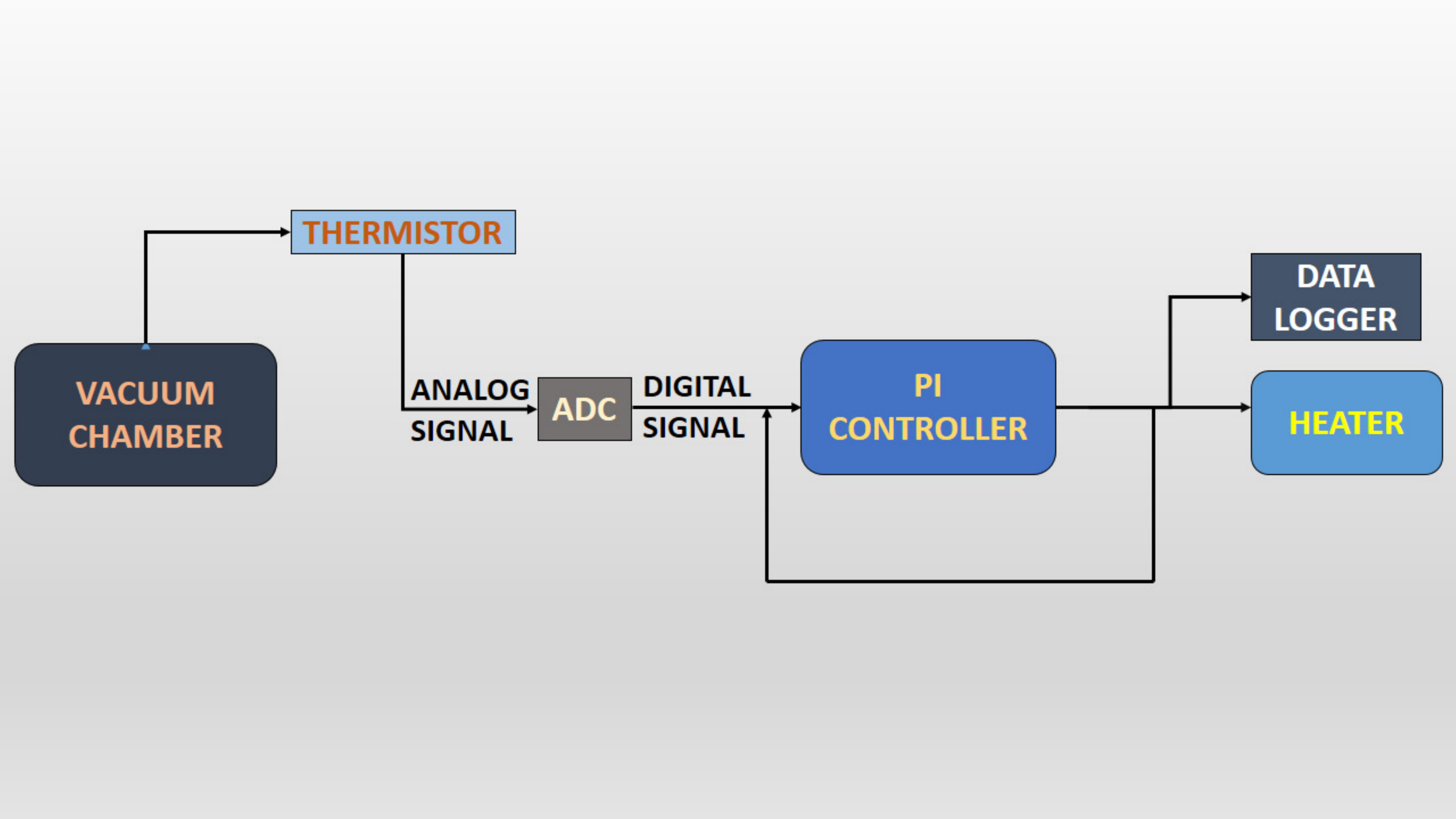}}
\caption{Temperature controller block diagram. The temperature of the chamber is sensed using thermistors and relayed to the Arduino board through an external 16 bit ADC for further processing.}
\label{fig:blk}
\end{figure}

In order to obtain stable reference lines using FP, minimum fluctuation in the temperature inside the vacuum enclosure has to be ensured. Arduino Uno microcontroller \cite{rf:uno} and off-the-shelf components have been used to design a digital Proportional-Integral (PI) temperature controller. Block diagram of control process is shown in Fig.~\ref{fig:blk}.

The nominal operating temperature for the enclosure was set 5~$^{\circ}${C} above the ambient with fluctuations not exceeding $\pm$0.05~$^{\circ}${C}. Four thermistors have been installed for temperature sensing; two inside wired through electrical feedthroughs and two outside the chamber. Printed polymer PTC heating elements, connected in series, are used to heat the chamber. The inner layer of thermal insulation is provided by nitrile rubber wrapped around the chamber tube. The entire setup is then kept inside a styrofoam box providing an outer layer of heat insulation. 

\begin{figure}[htbp]
\centering
\vbox{\includegraphics[width= 8cm, height = 8 cm,keepaspectratio]{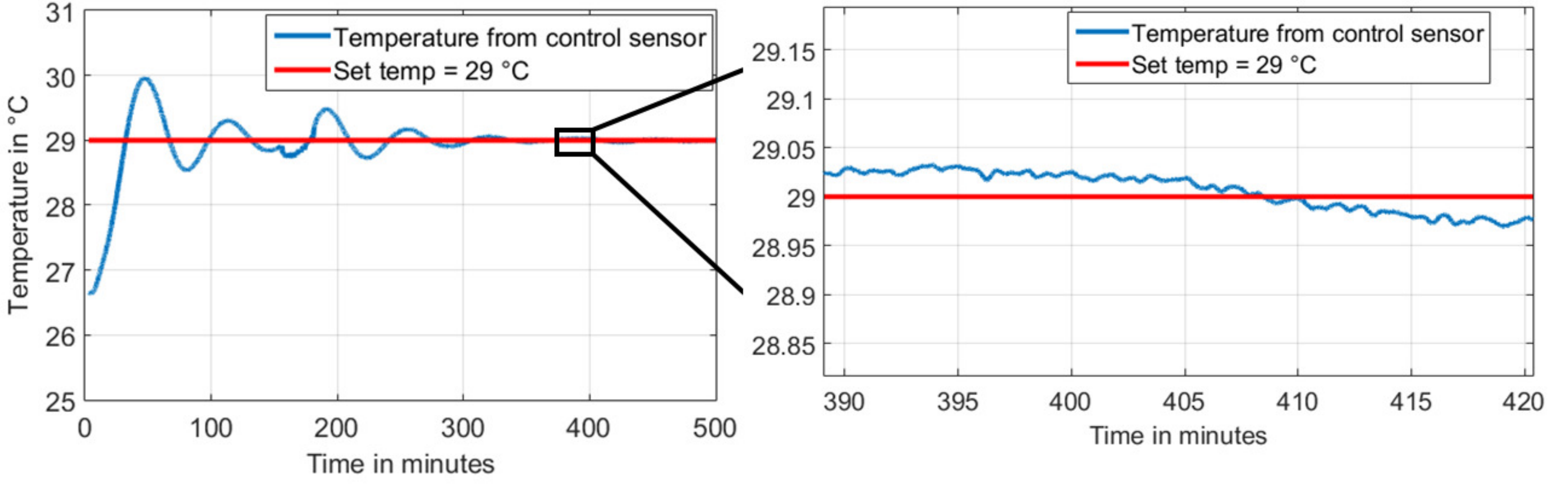}}
\caption{Left: Temperature of the FP chamber reaching a steady state after initial oscillatory cycle. Temperature from control sensor is given as input to the PI controller. Right: Zoomed in view showing the deviations within $\pm$0.05 $^{\circ}${C} from the temperature set-point during the steady state.}
\label{fig:tmp}
\end{figure}

The enclosure is initially heated in a controlled manner so that the temperature is ramped up to reach a set point, after which the feedback loop in the controller maintains the temperature with minimum excursions. Temperature stability of the setup has also been tested in the laboratory. The test was conducted for a period of 8.5 hours. The enclosure temperature is maintained with required stability as shown in Fig.~\ref{fig:tmp}. The maximum and minimum deviations in the temperature were within $\pm$0.05~$^{\circ}${C} during the test run.

\subsection{Optomechanical design}
\begin{figure}[htbp]
\centering
\vbox{\includegraphics[width=\linewidth]{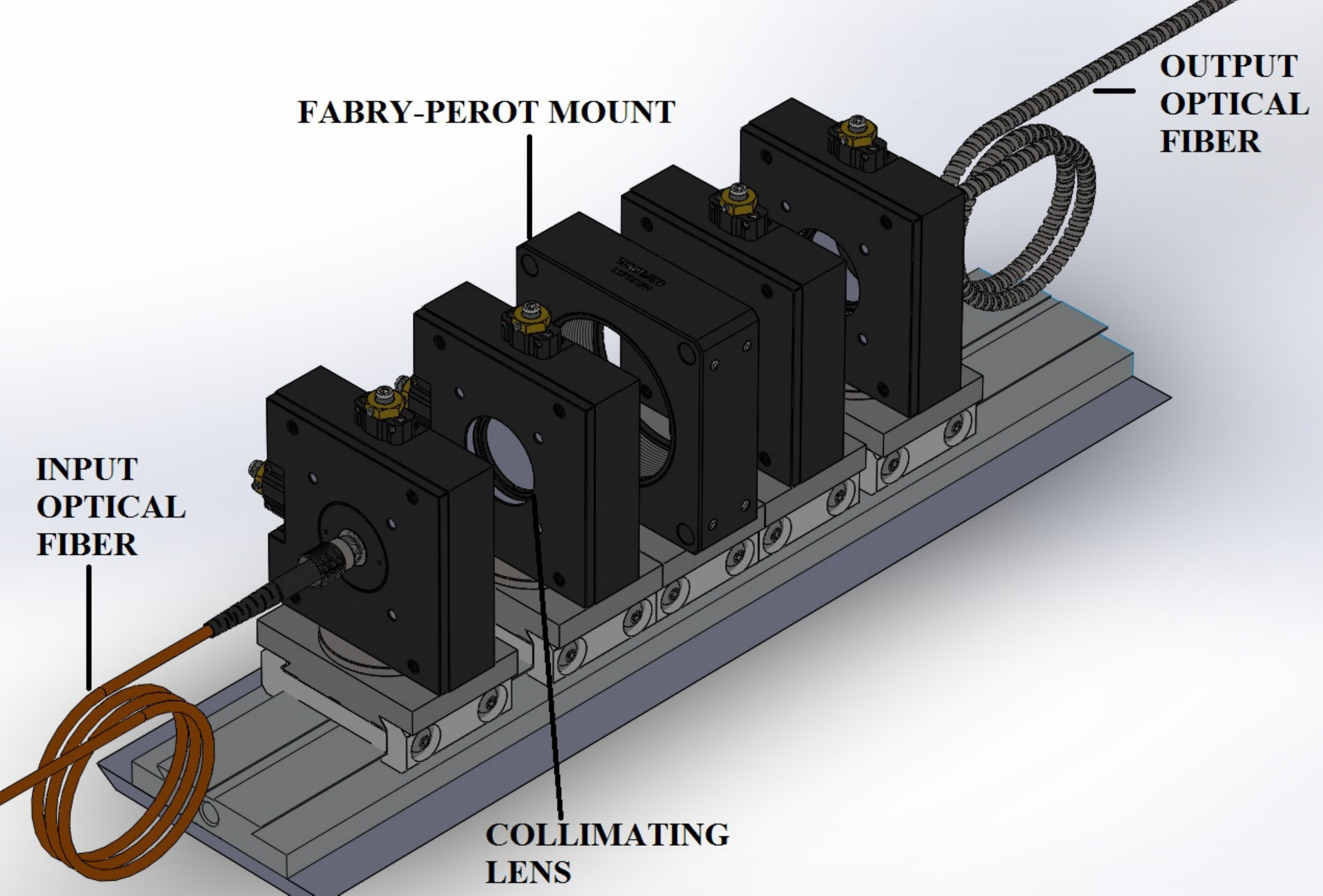}}
\caption{Schematic of the rail based mounting scheme for Fabry-P{\'e}rot.}
\label{fig:sldw}
\end{figure}

The optical components are being held in the vacuum chamber using a stable rail based mounting scheme as shown in Fig.~\ref{fig:sldw}.

An intensity stabilized broadband fiber coupled LED (Thorlab model MBB1F1) is used as the light source. Multimode fibers with core diameter 50~$\mu$m and 100~$\mu$m are used to carry light into and out of the chamber, respectively. Since the system is compact, achromat doublets (AC254-050-A-ML) of focal length 50~mm are used to collimate and collect light from the FP.

\begin{figure}[htbp]
\centering
\vbox{\includegraphics[width=\linewidth]{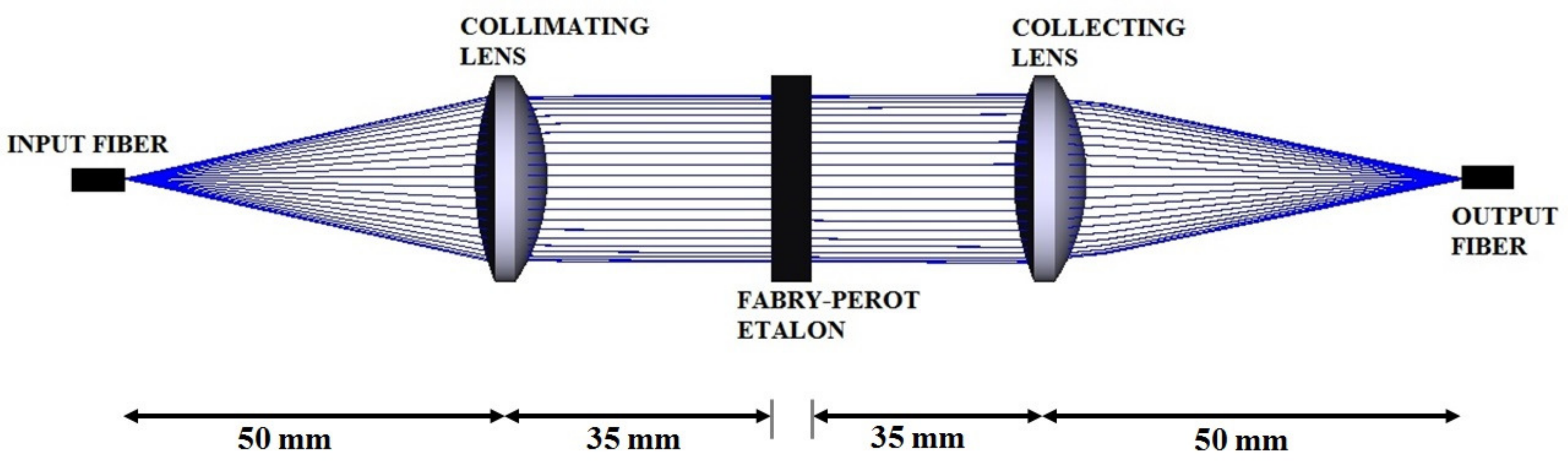}}
\caption{Ray diagram of the complete optical setup showing the beam path from input fiber to output fiber, through the lenses and the Fabry-P{\'e}rot.}
\label{fig:ray}
\end{figure}

We used a collimated beam of diameter 22~mm as input to the FP. The input beam slightly overfills the 20~mm  clear aperture of the FP. The output beam (dia =20~mm) from the FP was focused by an achromat doublet onto the 2nd fiber that carries it to the spectrograph calibration unit.  The end-to-end ray diagram of optical setup is shown in Fig.~\ref{fig:ray}.

\begin{figure}[htbp]
\centering
\vbox{\includegraphics[width=\linewidth]{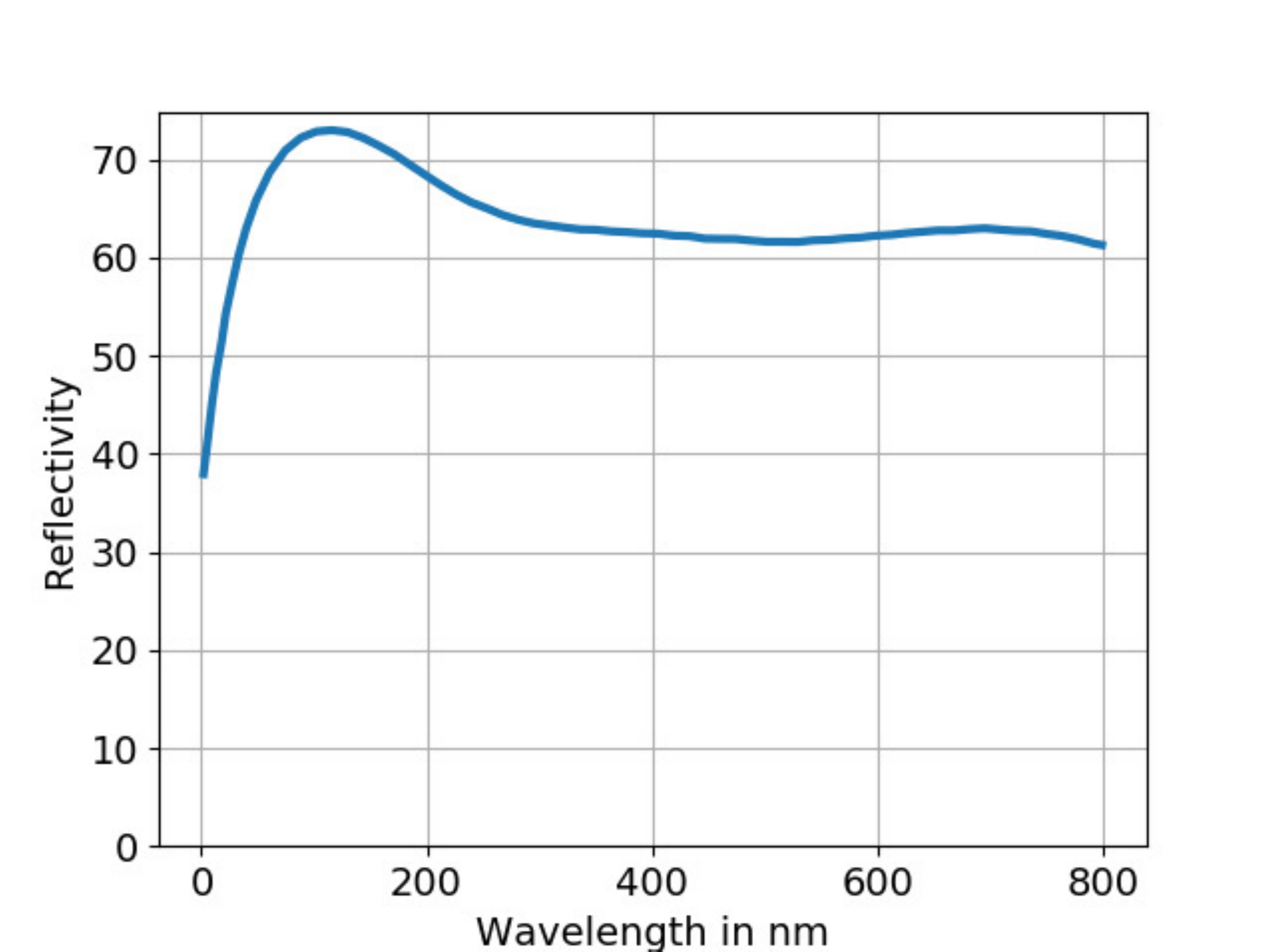}}
\caption{Reflectance spectra of the Fabry-P{\'e}rot cavity mirrors measured with monochromator. The spectra was provided by the manufacturer.}
\label{fig:reflect}
\end{figure}

The FP etalon was custom built by \textit {SLS Optics Ltd.,} to our specifications. The cavity mirrors were prealigned in the factory and securely sealed inside an aluminium holder. Both mirrors have broadband (500-750~nm) multi-layer coating. Fused silica was used as substrate and low thermal expansion Zerodur ring was used as spacer between two flat mirrors forming plane parallel cavity which is a preferred configuration for white-light optical filter. Reflectance spectra of the cavity mirrors is shown in Fig.\ref{fig:reflect}. Other specifications of the FP used in our setup are summarised in Table \ref{tab:prop}.

\begin{figure}[htbp]
\centering
\vbox{\includegraphics[width=\linewidth]{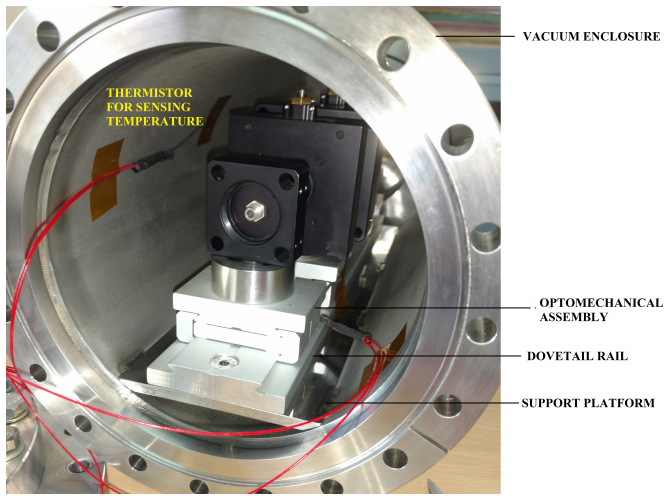}}
\caption{\textbf{FP optical assembly mounted on a support platform welded inside the vacuum chamber.}}
\label{fig:plt}
\end{figure}

Height of the optical setup including the dovetail rail is 115~mm and total length of the setup is 290~mm. A 5mm~thick stainless steel support platform is welded at the base of the chamber, on which the rail is mounted and secured with the help of countersunk screws. Figure~\ref{fig:plt} shows the optical setup mounted inside the chamber.

\begin{table}[htbp]
\centering
\caption{\bf Specifications of the Fabry-P{\'e}rot etalon}
\begin{tabular}{cc}
\hline
Etalon type & Air spaced  \\
Mirror material & Fused Silica  \\
Spectral coverage  & 500 – 750 nm \\
Cavity thickness & 5 mm$\pm$0.001 mm \\
Finesse & $\sim$6 \\
Clear aperture & 20~mm \\
Free Spectral Range & 30 GHz \\
Coating reflectivity & $\sim$60\%  \\
\hline
\end{tabular}
\label{tab:prop}
\end{table}

\section{Performance Test of Fabry-P{\'e}rot}
\label{sec:test}
\subsection{Test with Fourier Transform Spectrograph}
\begin{figure}[htbp]
\centering
\vbox{\includegraphics[width=\linewidth]{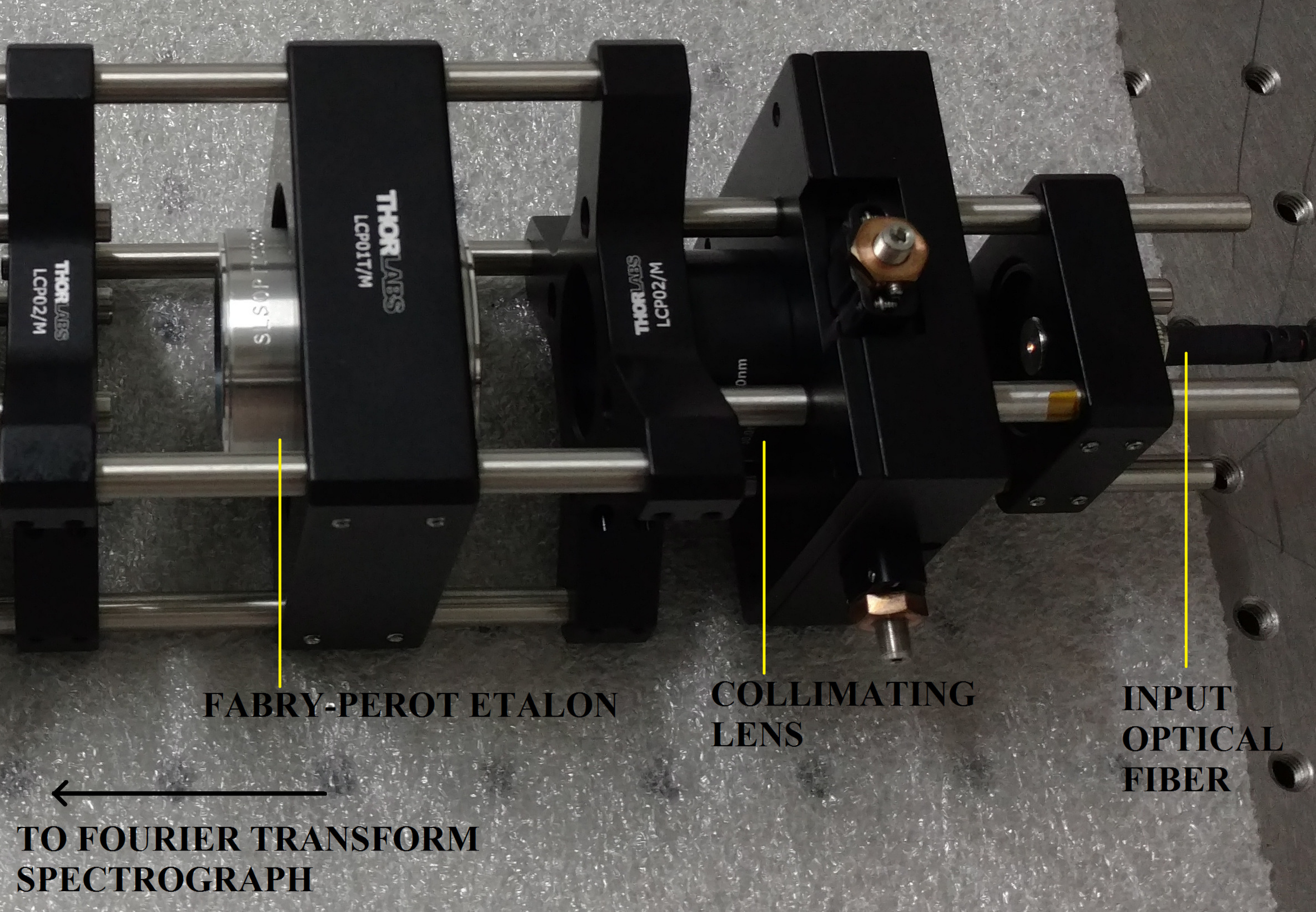}}
\caption{Cage based mounting scheme used for testing FP with FTS.}
\label{fig:fts}
\end{figure}

The high resolution Fourier Transform Spectrograph (FTS) facility at Bhabha Atomic Research Center (BARC), Mumbai, was used to test the etalon at normal temperature and pressure. Etalon was mounted on a cage assembly for the purpose of the test, as shown in Fig.~\ref{fig:fts}. The collimated output from the FP directly illuminates the FTS input port, to obtain transmission spectra at a resolution of 100,000.

\begin{figure}[htbp]
\centering
\vbox{\includegraphics[width=\linewidth]{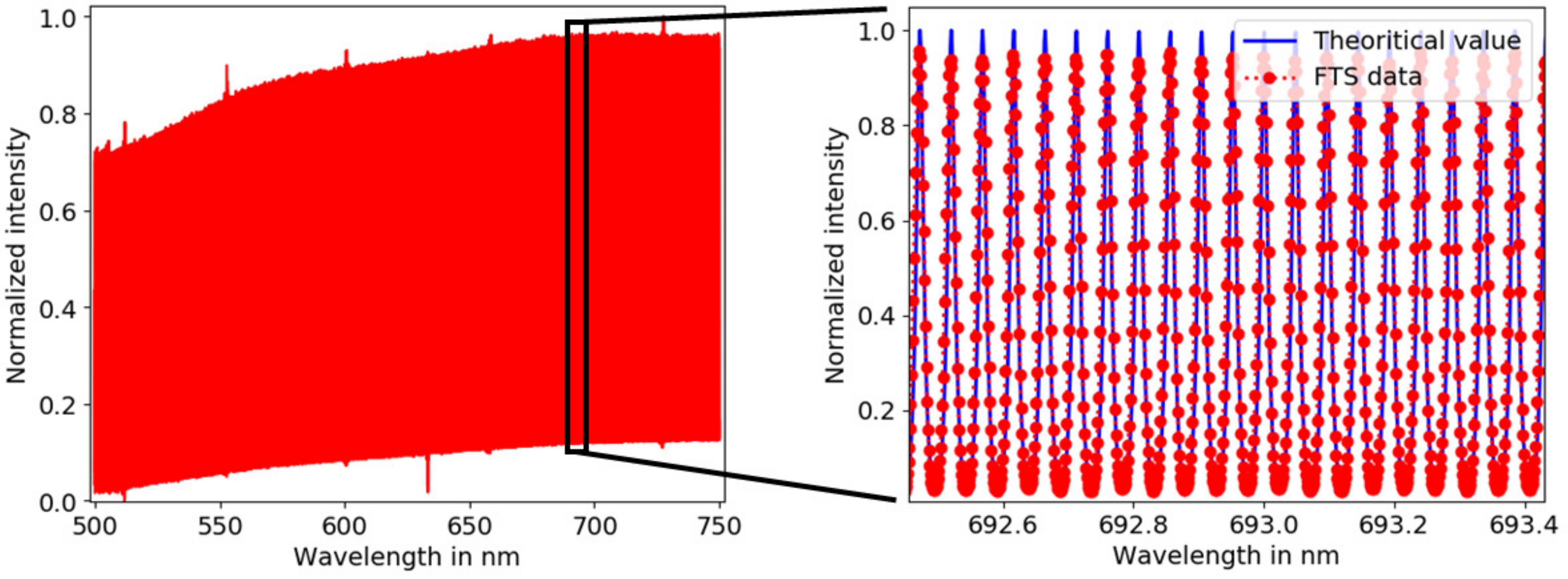}}
\caption{FP transmission spectra obtained from FTS with a broadband fiber coupled LED light source from Thorlabs (Model MBB1F1). Right: Narrow slice of FTS spectra, over plotted with theoretical spectra (blue curve)  computed from FP specifications provided by the vendor.}
\label{fig:zmc}
\end{figure}

FTS spectra of the FP has been analysed and plotted in Fig.\ref{fig:zmc}, along with theoretically generated FP spectra using the manufacturer provided parameters. From the analysis, the average cavity width of etalon was calculated as 4.999 mm as compared to 5mm$\pm$0.001 mm provided the manufacturer.

To investigate the impact of coating dispersion (see section 2\ref{sec:coating}) in  our FP, we looked for systematic variations in FSR. Other than random scatter, the measured FSR did not show any noticeable trend that can be attributed to coating dispersion. It is, therefore, likely that coating used in our FP had nearly a flat response. 

\subsection{FP installation on Hanle Echelle Spectrograph}
The etalon has been installed and tested with Hanle Echelle Spectrograph (HESP) on 2m Himalayan Chandra telescope at Indian Astronomical Observatory (IAO) located in Hanle, India \cite{rf:sriram}. The spectrograph design is based on white pupil concept and uses an image slicer in high resolution mode. It is housed inside a  thermally controlled isolated room below the observing floor. The spectra is imaged onto a 4k$\times$4k CCD with 15~$\mu$m pixel size. The thermal and mechanical stability along with dual fiber mode allows precision radial velocity measurements of stars. The specifications of the spectrograph are summarized in Table ~\ref{tab:hesp}.

\begin{table}[htbp]
\centering
\caption{\bf Specifications of Hanle Echelle Spectrograph (HESP) \cite{rf:sriram}.}
\begin{tabular}{cc}
\hline
Spectral Coverage & 350-1000~nm \\
Echelle grating & R2.1 \\
Resolution & 
\begin{tabular}{c}30000 (without image slicer) \\ 60000 (with image slicer) \end{tabular} \\
RV Accuracy & 20~m/s with dual fiber mode \\
Stability & 200~m/s (over a night) \\
Efficiency (includes telescope) & $\sim$22\% peak (at $\sim$600~nm) \\
\hline
\end{tabular}
\label{tab:hesp}
\end{table}

\begin{figure}[htbp]
\centering
\vbox{\includegraphics[width=\linewidth]{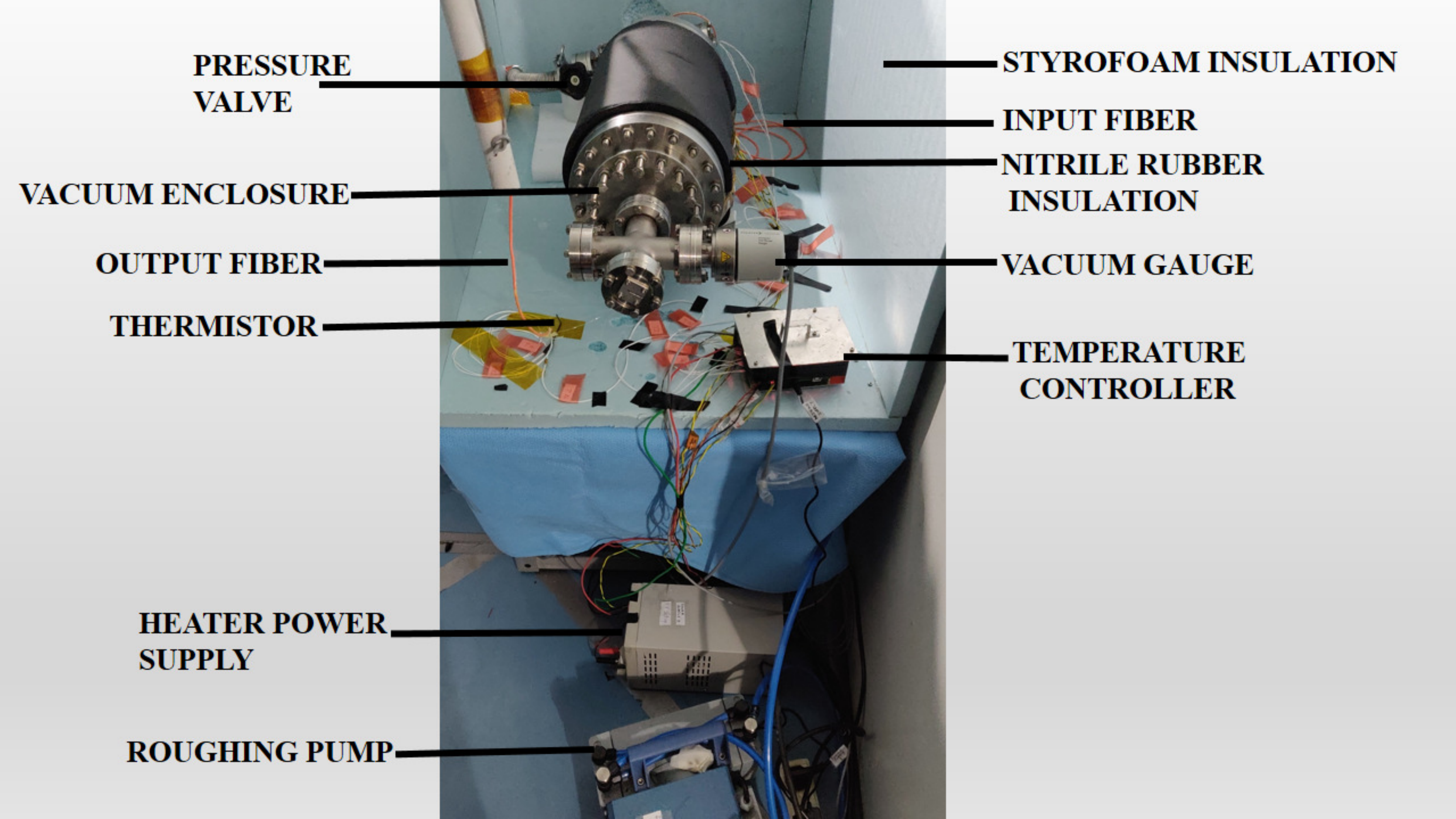}}
\caption{FP setup installed in the spectrograph room at Indian Astronomical Observatory, Hanle.}
\label{fig:hns}
\end{figure}

\begin{figure}[htbp]
\centering
\vbox{\includegraphics[width=\linewidth]{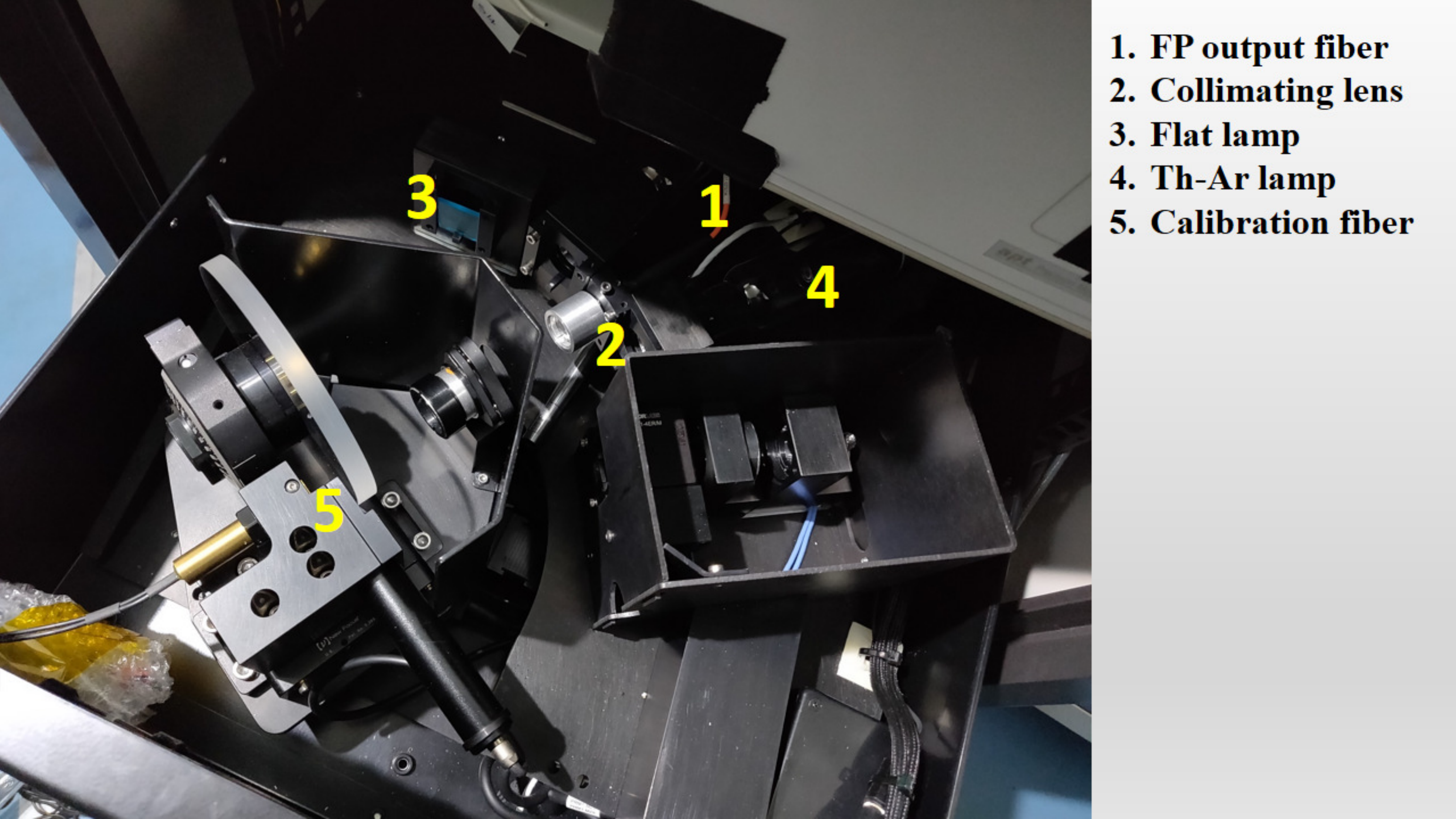}}
\caption{Calibration rack of HESP with the FP output fiber integrated into the rack.}
\label{fig:clb}
\end{figure}

The FP with associated optics is installed in a temperature controlled vacuum chamber with two layers of thermal insulation, as shown in Fig.~\ref{fig:hns}. The operating temperature for FP is 19$^{\circ}${C} and operating pressure is 0.025~mbar. The output from the FP is fed to calibration unit through a fiber of 100~$\mu$m size. The calibration unit holds the flat lamp, Th-Ar light source and a calibration fiber of 100 $\mu$m size that carries light into the spectrograph. The image of the calibration assembly is shown in Fig.~\ref{fig:clb}.

\begin{figure}[htbp]
\centering
\vbox{\includegraphics[width= 5 cm, height = 8cm, keepaspectratio]{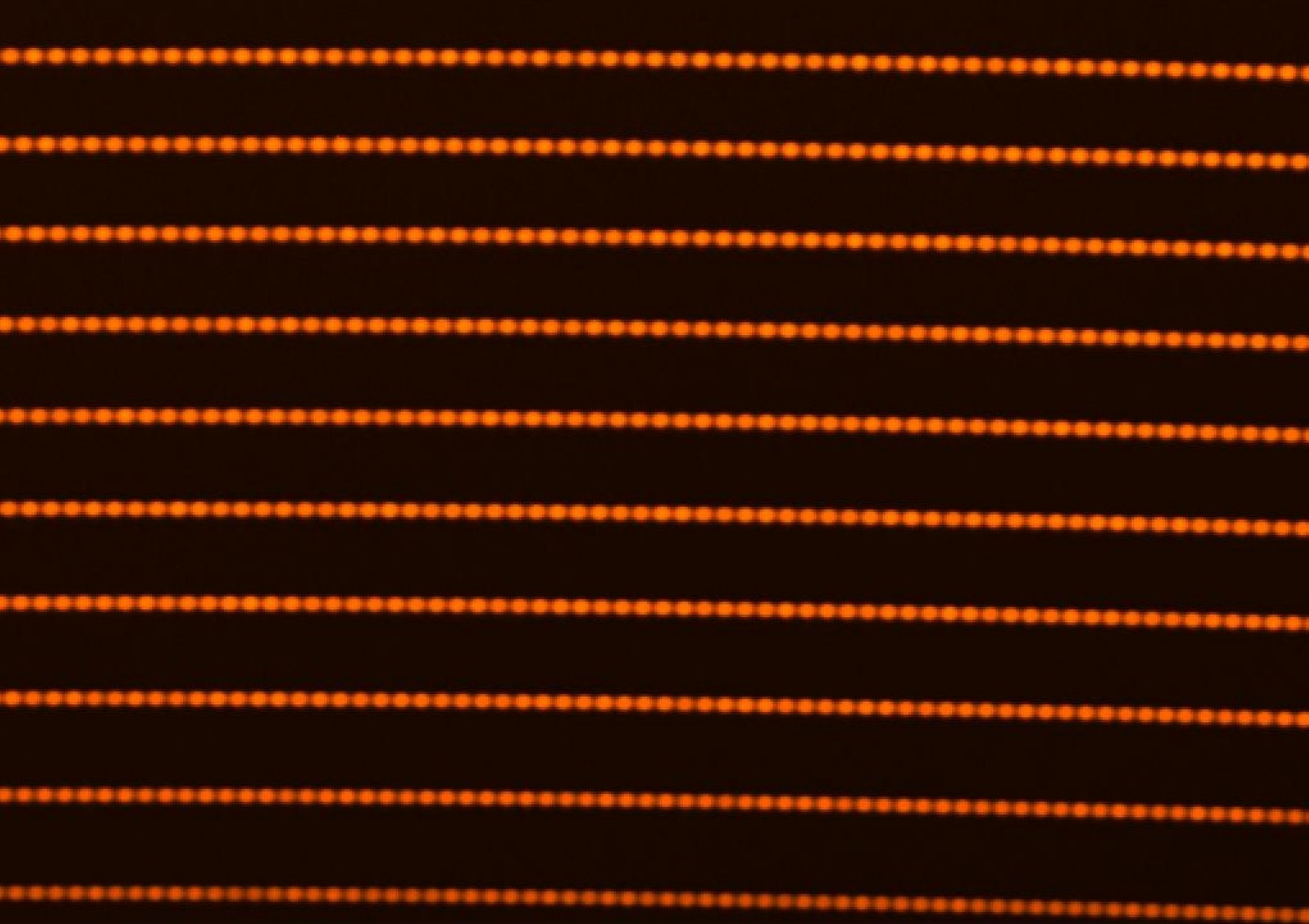}}
\caption{Fabry-P{\'e}rot spectra obtained from Hanle Echelle Spectrograph (HESP) at a resolution of R = 30,000, without using the image slicer.}
\label{fig:hsplow}
\end{figure}

\begin{figure}[htbp]
\centering
\vbox{\includegraphics[width=\linewidth]{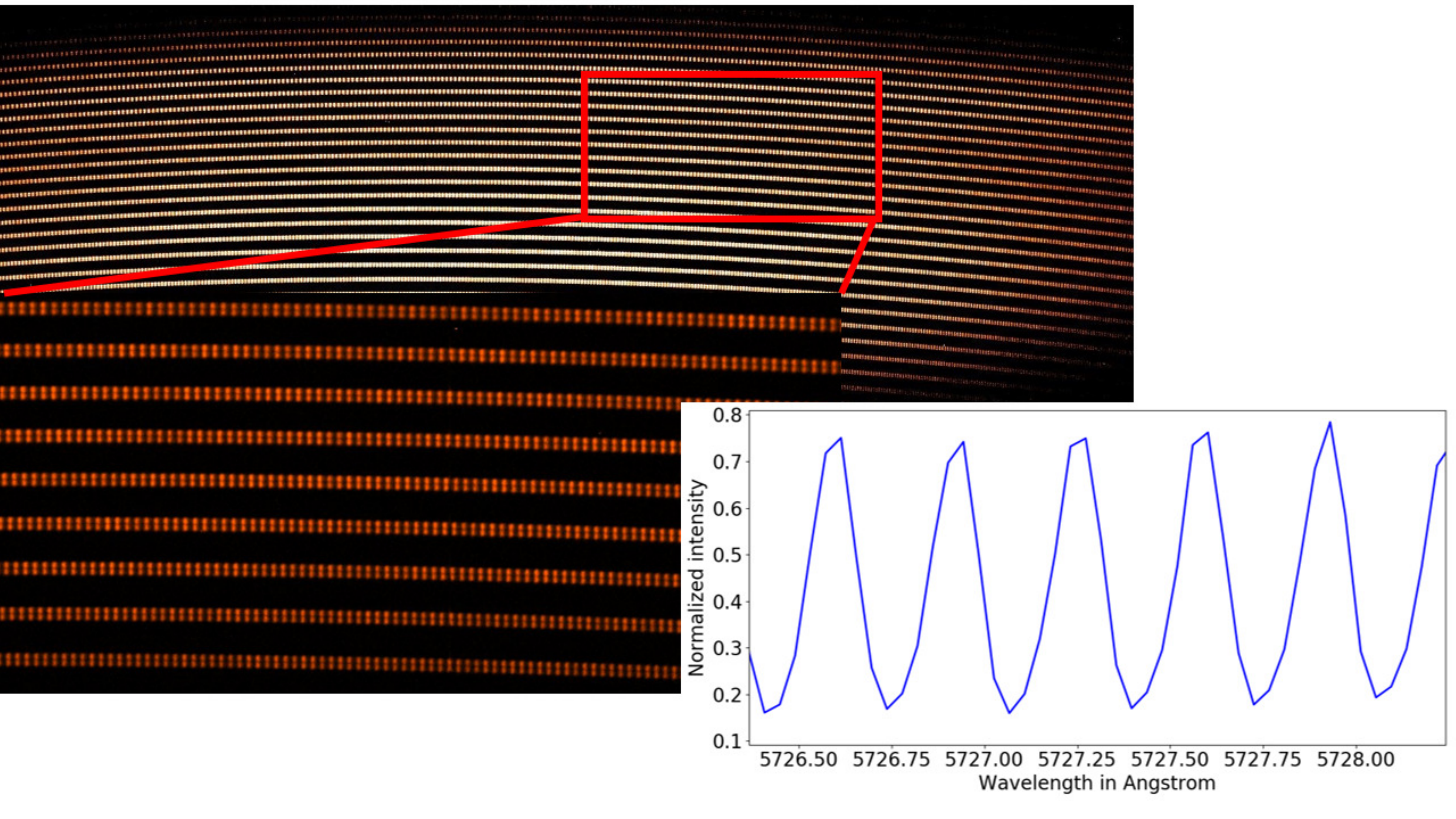}}
\caption{A high resolution (R=60,000) Fabry-P{\'e}rot spectra obtained from HESP. Top panel: A full CCD frame of the spectra. Bottom left panel: Zoomed in version focusing a part of the spectra.  Bottom right panel: Portion of an extracted spectral order showing the well resolved FP lines.}
\label{fig:spec}
\end{figure}

\begin{figure}[htbp]
\centering
\vbox{\includegraphics[width=\linewidth]{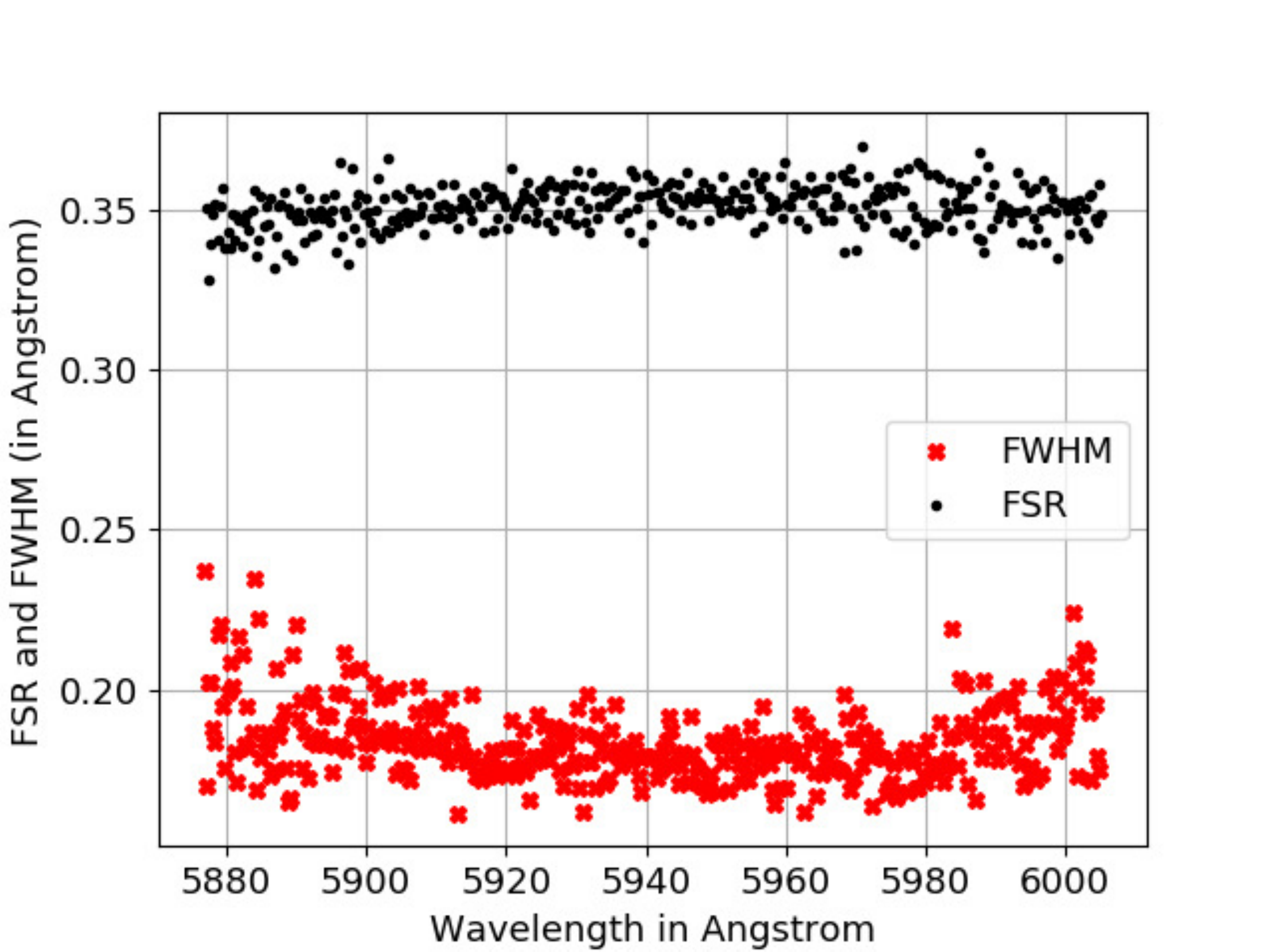}}
\caption{FSR and FWHM of the Fabry-P{\'e}rot lines as calculated from spectrograph data. Black dots indicate FSR and red crosses indicate FWHM in one of the orders in the spectra.}
\label{fig:ff}
\end{figure}

\begin{figure}[htbp]
\centering
\vbox{\includegraphics[width=\linewidth]{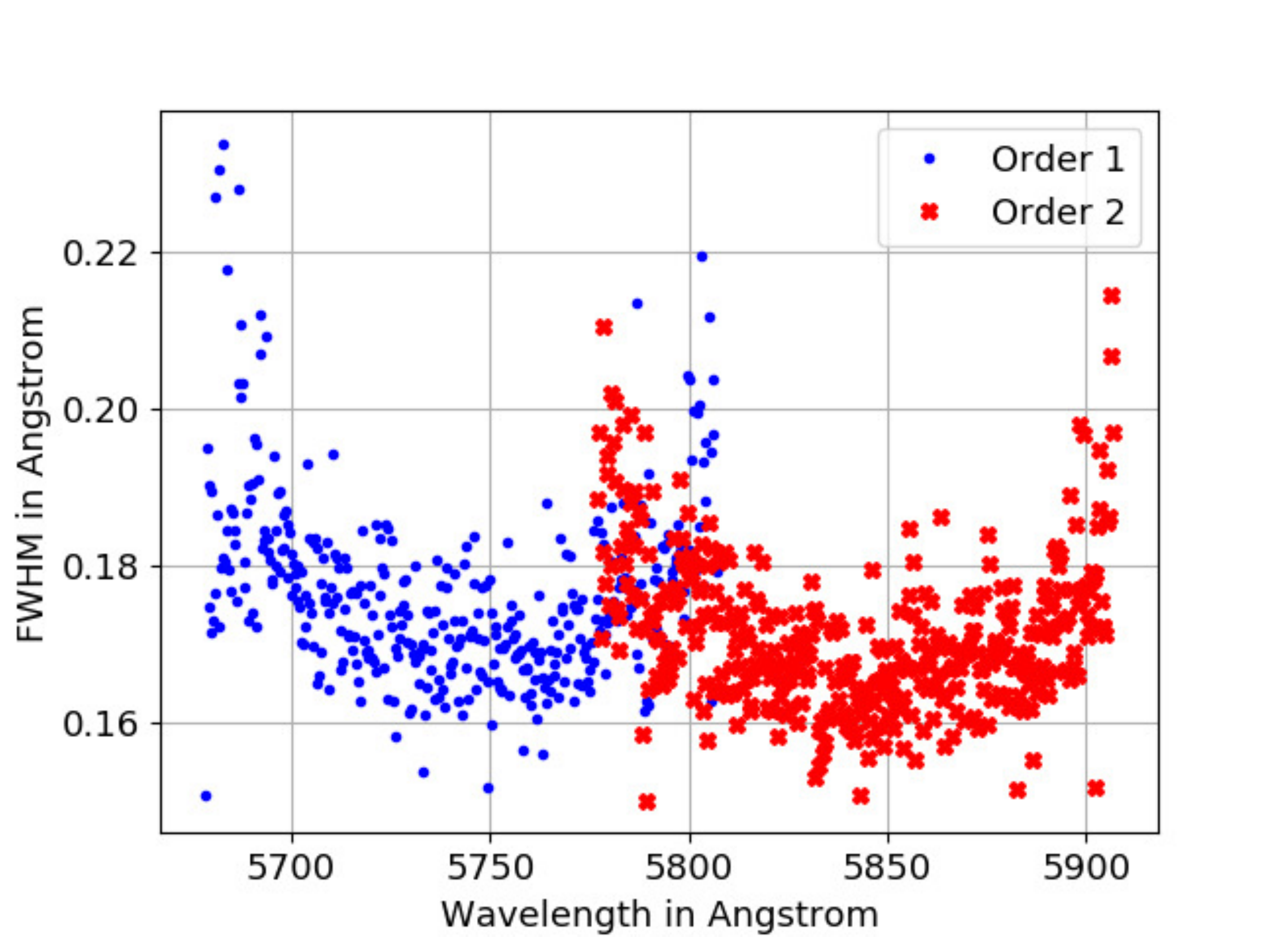}}
\caption{FWHM of Fabry-P{\'e}rot lines obtained in two adjacent orders showing distortion in FWHM at the edges. The overlapping is due to the wavelength overlap that commonly occurs at the edges of adjacent Echelle orders.}
\label{fig:fwhm}
\end{figure}

The spectra obtained from HESP, both in low resolution (R=30,000) and high resolution mode (R=60,000), is shown in Fig.~\ref{fig:hsplow} and Fig.~\ref{fig:spec}, respectively. FP lines were found across 24 orders of the spectrograph. Gaussian profile is fit to individual FP lines and the FSR and linewidth at each wavelength is determined from the data. Total number of FP lines detected are 6124 across the wavelength range 500-750~nm. Initial level of calibration of FP data has been done using Th-Ar to determine the wavelength of FP lines. A wavelength calibration model will be developed using the techniques employed in \cite{rf:bauer,rf:newwav}. Free Spectral Range (FSR) for a spectropgraph is the wavelength range in a spectral order without overlap from the adjacent orders whereas FSR for FP is the distance between consecutive transmission peaks. Figure\ref{fig:ff} shows the FSR and FWHM values of FP plotted as a function of  wavelength, for one of the orders. Preliminary analysis of FP calibration frames show that the spectrograph has a characteristic presence of field curvature and distortion introduced by the optics. We have observed a curvature in FWHM in all the orders of the spectra. Apart from this, the instrumental artifacts are also seen at the edges of each order as shown in Fig.~\ref{fig:fwhm}. This opens up new possibilities of spatially mapping the spectrograph point-spread function and other artifacts from the FP spots. Incorporating such input would help us developing a better wavelength solution for the spectrograph. A detailed model of spectrograph aberrations mapped from FP data with improved wavelength solution will be presented in a follow-up work under study. 

\section{Summary}
\label{sec:summary}
We have developed a low-cost, passively stabilized Fabry-Perot based wavelength calibrator for Hanle Echelle Spectrograph installed on 2m class HCT. Off-the-shelf components were used for the construction of the setup with the temperature controller and fiber feedthrough developed in-house. The FP was first tested using a FTS and later integrated with the wavelength calibration unit of the HESP. Some initial test runs have been conducted with the spectrograph. Table~\ref{tab:comp} shows the comparison among values for finesse and cavity width as obtained from the test data. The observed difference can be attributed to finite size of the optical fiber and also the degradation of optical coating of the FP since its purchase over 3 years back.

\begin{table}[htbp]
\centering
\caption{\bf Comparison of FP parameters obtained from datasheet, FTS and HESP.}
\begin{tabular}{cccc}
\hline
FP Parameter & FP datasheet & FTS test & HESP data \\
\hline
Cavity Width & 5 mm & 4.999 mm & 4.983 mm \\
Finesse & 6 & 2.73 & 2 \\
\hline
\end{tabular}
\label{tab:comp}
\end{table}

For better accessibility and on-site alignment, some simplification of the setup are necessary. We also plan to use a 50~$\mu$m fiber as input to limit FP line broadening resulting from the extended source size. With these changes in place, we would be conducting on-sky tests using standard radial velocity target stars to evaluate the RV performance and then commission the instrument as an add-on calibration module for HESP to facilitate regular science observation. Aging of etalon spacers and coating degradation over time can cause intractable frequency drifts in a passively stabilized etalon \cite{rf:dual}. Therefore, we are also exploring the possibility of actively tracking the FP cavity with a frequency stabilized diode laser \cite{rf:reiners}. 

\noindent\textbf{Funding.} This project is supported by Science and Engineering Research Board (SERB), Department of Science and Technology (DST), India, under grant no. EMR/2014/000941.

\noindent\textbf{Acknowledgements.} The authors want to acknowledge Dr. M. N. Deo and Dr. H. Bhatt for allowing the use of high-resolution FTS facility for characterizing the Fabry-P{\'e}rot etalon at Bhaba Atomic Research Center, Mumbai. The authors also thank Mr. S. Sriram, Mr. S. Kathiravan, Mr. A. Kumar and Ms. Devika D. for technical support. 

\medskip
\noindent\textbf{Disclosures.} The authors declare no conflicts of interest.
\medskip

\bibliography{reference}

\bibliographyfullrefs{reference}

\end{document}